# Evaluating reliability of complex systems for Predictive maintenance

**Dongjin Lee**
**Arizona State University**
**School of Computing, Informatics, and Decision Systems Engineering**
**699 S.Mill Avenue, Tempe, AZ 85281, USA**

**Rong Pan**
**Arizona State University**
**School of Computing, Informatics, and Decision Systems Engineering**
**699 S.Mill Avenue, Tempe, AZ 85281, USA**

## Abstract

Predictive Maintenance (PdM) can only be implemented when the online knowledge of system condition is available, and this has become available with deployment of on-equipment sensors. To date, most studies on predicting the remaining useful lifetime of a system have been focusing on either single-component systems or systems with deterministic reliability structures. This assumption is not applicable on some realistic problems, where there exist uncertainties in reliability structures of complex systems. In this paper, a PdM scheme is developed by employing a Discrete Time Markov Chain (DTMC) for forecasting the health of monitored components and a Bayesian Network (BN) for modeling the multi-component system reliability. Therefore, probabilistic inferences on both the system and its components' status can be made and PdM can be scheduled on both levels.

## 1. Introduction

Scheduling maintenance at right time and right place is very important for keeping a system's reliability at the required level. Many studies have been conducted for improving maintenance scheduling to achieve more efficient and economical use of assets that can increase equipment lifetime, decrease operating cost, and produce high-quality products. Conventional maintenance strategies adopted in the current industry invoke maintenance actions based on system's work hours. Any unexpected failure may cause significant production loss and reduce system efficiency. Even Preventive Maintenance (PM) that schedules maintenance based on historical repair data of the system cannot be effective in preventing these unexpected failures, as PM does not use current system and components' health to make system reliability prediction.

This research proposes an approach to Predictive Maintenance (PdM) for complex systems by using an integrated framework of Discrete Time Markov Chain (DTMC) and Bayesian Network (BN). A complex system is defined as a system consisting of multiple components and the dependency between system reliability and component reliability may not be completely known. We propose to use BN for system reliability modeling so that the uncertainty in reliability modeling can be explicitly taken into consideration; in the meantime, the evolution of components' health states will be modeled by DTMC. A main purpose of PdM is to carry out maintenance activities when they are indeed needed. PdM becomes possible only when a system's future health state can be forecasted by using data collected from on-equipment sensors. In this paper, we will focus on the PdM of multi-level hierarchical systems, where a system consists of subsystems, and a subsystem consists of components. Such systems widely exist in various fields [1]. Sensors will be put on these components to monitor their health states, while system reliability needs to be estimated based on the information from sensors.

## 2. Literature Review

System degradation has been studied by many researchers. Most of them focused on the systems that has a single component or the system that has multiple components with well-defined system reliability function. A Bayesian approach for a degradation model is proposed for a single component to develop a residual lifetime distribution in [2]. Another Bayesian approach is suggested in [3], where parameters in the degradation model in [2] have a joint distribution. The suggested degradation model in [2] is extended to the maintenance strategy in [4]. Ref. [5] analyzed multi-component systems consist of identical elements using dynamic programming for a purpose of preventive maintenance. This study is generalized by [6] to the system consisting of non-identical elements. Ref. [7] provided a method to model a non-repairable single component and a repairable multi-component systems using Markov process and Monte Carlo simulations where the system is continuously monitored and the maintenance schedule is optimized. Genetic Algorithm is applied to [7] for finding the optimal thresholds in [8]. Bayesian networks (BNs) and dynamic Bayesian networks (DBNs) also have been used for modeling multi-component systems. Ref. [9, 10] used DBNs for modeling multi-component systems and maintenance purposes. Both works show differences on approaching to the problems from perspectives of defining nodes of networks. [11, 12] applied BNs to model multi-component and multi-stage systems. Ref. [13] used BNs to analyze multilevel system where information comes from different levels of the system.

There are several papers focusing on the relationship of system and component reliability in a multi-component system. Ref. [14] proposed a general modeling approach of multi-component systems using multiple degradation paths. Ref. [15] developed an approach where components share common environment, which makes correlation between components. A stochastic methodology was proposed in [16] when interactions between degradations of components exist. Ref. [1] proposed a BN approach for modeling multi-level systems considering component dependency where each node has binary state, and ref. [17] extended it to components with multi-state. Ref. [18] used BNs and Bayesian inference for analyzing multi-component systems considering dependency of components. Ref. [19] proposed BNs for reliability of large structures considering correlation of components. A general approach for optimal maintenance considering dependency of components in multi-components system is provided in [20]. In this paper, we propose a framework of integrating BN and DTMC for predicting a multi-component system's health state, with the consideration of uncertainty in system reliability modeling.

## 3. Methodology

### 3.1 Discrete Time Markov Chain

A DTMC is a stochastic process $\{X_n, n > 0\}$ in a discrete state and time space ruled by the Markov property. The Markov property is satisfied if future events are independent of the past and only depend on the current state.

$$P(X_{n+1} = x | X_n = x_n, X_{n-1} = x_{n-1}, \ldots, X_1 = x_1) = P(X_{n+1} = x | X_n = x_n) \qquad (1)$$

where $X_n = x_n$ indicates that the random variable $X$ takes state $x_n$ at time $n$.

Let $p_{ij} > 0$ represent the one-step transition probability from the current state $i$ to the next state $j$. The transition probabilities between all states form the one-step Transition Probability Matrix (TPM), $P$.

$$P = \begin{bmatrix} p_{00} & p_{01} & \cdots & p_{0N} \\ p_{10} & p_{11} & & p_{1N} \\ \vdots & & \ddots & \vdots \\ p_{N0} & p_{N1} & \cdots & p_{NN} \end{bmatrix} \qquad (2)$$

The $n$-step TPM can be calculated by the power function of one-step TPM, i.e. $P^n$, where its element, $p_{ij}^{(n)}$, denotes the probability of transitioning from state $i$ to state $j$ over $n$ steps.

### 3.2 Bayesian networks

A BN is a directed acyclic graph (DAG) where nodes represent random variables and edges indicate direct probabilistic dependencies between the nodes [21], and joint probability distributions are represented using conditional independence. Thus, the BN graphically represents probabilistic causal relationships between the random variables and directions of information flow [19]. In a BN with nodes $\{X_1, X_2, \ldots X_n\}$ and edges between the nodes, $X_i$ is a parent of $X_j$ if there is an arrow from $X_i$ to $X_j$, and $X_j$ is a child of $X_i$. A node without parents is called a root node, and a node without child is called a leaf node. Figure 1 is a simple BN example. In this example, $X_2$ and $X_3$ are root nodes and they are also the parents of $X_1$, while $X_1$ is a leaf node and the child of $X_2$ and $X_3$.

In order to build a BN, the topology of BN has to be structured first. Then, the BN is parameterized by quantifying the relationships between nodes. Parameters are conditional probabilities for each node given the values of its parents, and listed in a conditional probability table (CPT). Because root nodes do not have parents, CPTs contain marginal probabilities. Table 1 and Table 2 show CPTs for the BN in Figure 1.

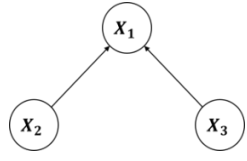

Figure 1: A BN example

The conditional independence of BN implies the chain rule that is useful to calculate the joint probability. The full joint distribution is factorized using the chain rule according to the topology of BN. For example,

$$P(X_1, X_2, \ldots, X_n) = \prod_{i=1}^{n} P(X_i | pa(X_i)) \tag{3}$$

where $pa(X_i)$ is the set of parent of node $X_i$.

In the BN, a marginal distribution for each node can be calculated using the law of total probability. For example, the marginal probability of node $X_i$, $P(X_i = x)$ is

$$P(X_i = x) = \sum_{c \in C} P(X_i = x | pa(X_i) = c) \prod_{pa_i(X_i) \in pa(X_i)} P(pa_j(X_i) = c_k) \tag{4}$$

where $C$ is a set of all combinations of the parent of $X_i$, $c$ is an element of $C$ in a vector form, $pa_j(X_i)$ is $j^{th}$ parent of $X_i$, and $c_k$ is an element of $c$ that is a value of the corresponding parent.

If conditional probabilities in the CPT of a child node are listed as vectors, Eq. (4) can be represented in a more compact way. First, we define the CPT of a child node as matrix $T$, and $T_{(i)}$ denotes the column vector corresponding to the child node value $i$. Define a vector $V$ as the vectorized Kronecker product of marginal distributions of parent nodes, where the Kronecker product is defined as

$$A \otimes B = \begin{bmatrix} a_1 B \\ \vdots \\ a_m B \end{bmatrix} \tag{5}$$

where $A$ is an $m \times 1$ vector and $B$ is a $p \times 1$ vector. Then, the marginal probability of child node can be expressed as

$$P(X = i) = T_{(i)}^t V \tag{6}$$

where $T_{(i)}^t$ is the transpose of $T_{(i)}$. For example, the CPT of child node in Table 1 is a $4 \times 2$ matrix $T$, and the Kronecker product is applied to the parent node distributions. We have $V = [p_2 p_3, p_2(1-p_3), (1-p_2)p_3, (1-p_2)(1-p_3)]^t$. Then, $P(X_1 = 0)$ is given by

$$P(X_1 = 0) = T_{(0)}^t V \tag{7}$$

Table 1: Conditional probability table

|  |  | $X_3 = 0$ | $X_3 = 1$ |
|---|---|---|---|
| $X_2 = 0$ | $X_3 = 0$ | $p_{00}$ | $1 - p_{00}$ |
| $X_2 = 0$ | $X_3 = 1$ | $p_{01}$ | $1 - p_{01}$ |
| $X_2 = 1$ | $X_3 = 0$ | $p_{10}$ | $1 - p_{10}$ |
| $X_2 = 1$ | $X_3 = 1$ | $p_{11}$ | $1 - p_{11}$ |

Table 2: Marginal probability tables for the root nodes

| $X_2 = 0$ | $X_2 = 1$ |
|---|---|
| $p_2$ | $1 - p_2$ |

| $X_3 = 0$ | $X_3 = 1$ |
|---|---|
| $p_3$ | $1 - p_3$ |

Using BN to model system reliability allows the uncertainties of the relationship of system and component reliability in a complex system to be explicitly considered in the model. Ref. [23] compared BNs with block diagrams or fault trees as the graphical means for modeling system reliability. Ref, [18] discussed how to estimate the conditional probabilities of a BN using aggregated system-level and component-level data, and ref. [24] extended the discussion to the scenario of having simultaneous failure observations on both system and component levels. In this paper, we assume that the CPTs are known, possibly estimated from the system design with historical repair data.

### 3.3 Proposed method for Predictive maintenance

To perform PdM, we need to forecast the future system reliability with current sensor data so that we can make a decision whether or not a maintenance action is to be taken. To accomplish this goal, some assumptions are made:

   i.    The on-equipment sensors provide true health states of components.
   ii.   The health state cannot be improved without maintenance.
   iii.  Nodes in lower levels directly or indirectly influence nodes in higher levels but not the opposite direction.

The system reliability is a function of component reliability [22], and it is a marginal probability of a node representing the system on a BN. For example, the BN in Figure 1 represents a system, $X_1$, consisting of two components, $X_2$ and $X_3$. If all nodes are binary random variable with functional state 0 and dysfunctional state 1, Eq. (7) is the system reliability. Thus, to estimate the future system reliability, future components' health states are required. In this study, transition of component's health state is assumed as a stochastic process, and future state only depend on the current state. Thus, a DTMC model is employed for the transition of component's health state.

Suppose a system consists of $N$ components, and the $i^{th}$ component has discretized health states $\{0, 1, \ldots, f_i\}$ where 0 is the healthy state, $f_i$ is the failure state, and any states in between are degradation states. Then, each component has a $(f_i + 1) \times (f_i + 1)$ one-step TPM.

$$P_i = \begin{bmatrix} p_{00} & p_{01} & \cdots & P_{0f_i} \\ 0 & p_{11} & & p_{1f_i} \\ \vdots & & \ddots & \vdots \\ 0 & 0 & \cdots & 1 \end{bmatrix} \quad (8)$$

Note that $p_{ij} = 0$ if $i > j$ according to the second assumption at the beginning of section 3.3.

Given the current component's health state provided by the on-equipment sensors, the probability that the component's state will become any specific state after $n$-step transitions can be found as elements in $P_i^n$. Any row of $P_i^n$ plays a role of a marginal distribution for the corresponding component when its initial state is fixed. Then, based on the marginal distributions for components, Eq. (4) or Eq. (6) can be used to estimate the future system reliability.

In a general case, a hierarchical system consists of $N$ components with $l$ levels where each component, subsystem, and system have multi-state as in Figure 2. $X_{i,j}$ indicates the $j^{th}$ node in the $i^{th}$ level. Every component in level $l$ has a TPM. Suppose the on-equipment sensors give a vector of current components' health state $h = \{h_1, h_2, \ldots, h_N\}$, where $h_i$ is the health state of $i^{th}$ component, and we want to forecast the system reliability after the $n$-step. A marginal distribution that will be used as CPT for each component is the $h_i^{th}$ row of $n$-step TPM of $i^{th}$ component. The marginal distributions of all sub-system in $l - 1$ level are calculated using the CPTs of level $l$, and the same scheme is repeated until the marginal distribution of the node representing the system in the first level is achieved.

$$P(X_{i,j} = k) = T_{i,j,(k)}^t V_{i+1} \quad (9)$$

where $T_{i,j,(k)}$ and $V_{i+1}$ indicate the column of state $k$ in a CPT of the $j^{th}$ node at level $i$ and the vector calculated by the Kronecker product using marginal probabilities of corresponding parents at level $i + 1$, respectively.

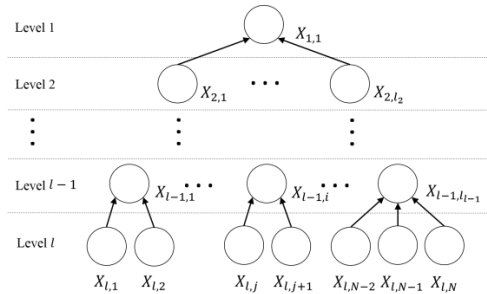

Figure 2: Multi-Level Hierarchical System

## 4. Simulation

With given TPMs and CPTs, we ran Monte Carlo simulation on a hypothetical system, which consists of two subsystems (subsystem 1 and subsystem 2), and the subsystem 1 (subsystem 2) is composed of component 1 and component 2 (component 3 and component 4). The system has two states (functional state is 0 and dysfunctional state is 1), the subsystem 1 (sub-system 2) has seven states (six states), the component 1 and component 3 has seven states, and the component 2 and component 4 has six states, where 0 indicates the best state for all nodes.

The results are shown in Figure 3. Four scenarios are made based on different current health states of components. The bold line indicates the forecasted future system reliability over the next one hundred time stamps, and other colored lines denote the realizations of future reliability. The results show that the forecast is more precise as we predict the reliability of closer future, or far enough in future so the system will certainly fail. The forecast becomes inaccurate when we estimate the reliability in the middle time frame. This is because the number of combinations of components' health states is gradually increasing as the system continues operation, but after a while it will steadily decrease because all components' health states eventually arrive at the failure state. Another characteristic is that the forecast is more precise as current component states are more degraded. It is because, as a degraded state cannot transition to a healthier state, the number of combinations of components' health states in the future is reduced.

Maintenance decisions are made for two scenarios. Assume the current state of components are (0, 0, 0, 0) (sencaario 1) and (2, 1, 2, 1) (scenario 2), under an assumption that the system is working until maintenance. For the first scenario, system reliability threshold are set to be 0.9 and the current system reliabilities are 0.99. the forecasted system reliabities at time 13 and 14 are 0.9 and 0.89, respectively; thus, it is necessary to schedule a future maintenance action immediately after the time point 13. In addition, component's health can be predicted using its marginal distribution at a given future time point. For example, it can be predicted that the component 1 is the most deteriorated component at time 13, based on the prediction of expected health states of all components at that time; thus, the maintence crew should prepare to repair component 1. Moreover, as these comonponts are continuously monitored by sensors, we can keep updating their health states. For example, if after 2 time lags the actual health states of components become as what in sencario 2, we should update our maintenance scheduling based on the latest inforamtion. Under the scenario 2, the forecasted system reliabilities at time 8 and 9 will be 0.9 and 0.88, respectively. Therefore, the maintenance action should be scheduled after time 8 (it is ahead of the previous schedule). Moreover, the components that are expected to be repaired have become components 1 and 3.

In case when both components and subsystems are monitored by sensors, in order to estimate the system reliability, we do not need the information of components if sensor data of the subsystem consisting of the component is available. However, for maintanance, the health state of any component that is not directly monitored can be inferred from the state of subsystem.

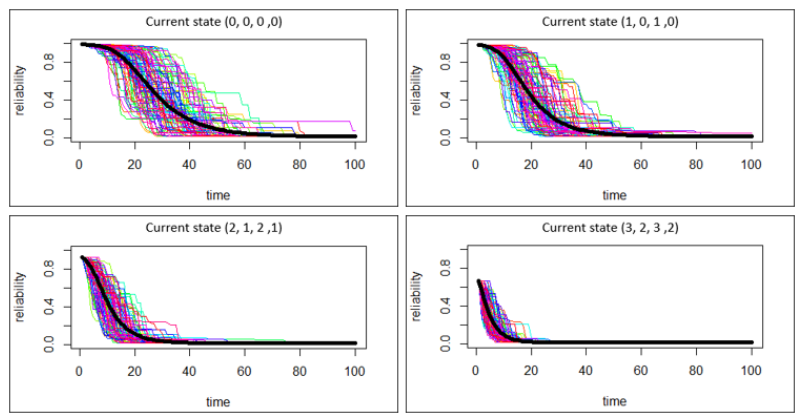

Figure 3: Monte Carlo simulation results

## 5. Conclusion

In this paper, we proposed a system reliability forecasting framework for implementing predictive maintenance. Our method can be applied on complex systems where dependencies among component, subsystem and system are not deterministic. The main objective of this framework is to forecast the future system reliability, which is followed by a maintenance decision. In order to consider the change of components' health states over time, DTMC is employed. DTMC and BN are connected through transition probability matrixes, and the system reliability is predicted by using components' future health states on the BN.

One of limitations of our approach is the use of the Markov property. Holding the Markovian assumption, we can simplify the problem and improve the computational efficiency. However, ignoring the deterioration history of components may lead to large error in prediction. Thus, generalizing the Markovian assumption could increase the accuracy of forecast. Estimating TPMs and CPTs which cannot be trivial in practice can also be another limitation. These two limitations will be investigated in our future research.